\documentclass[twocolumn,aps,prb,epsf,showpacs]{revtex4}
\usepackage{amsmath}
\usepackage{epsfig}
\usepackage{epsf}
\usepackage{array}
\usepackage{color}
\usepackage{graphicx}
\usepackage{epstopdf}
\begin{document}

\title{Fluctuation-driven first order behavior near the $T=0$ two dimensional stripe to fermi liquid transition }
\author{A. J. Millis}
\affiliation{
Department of Physics, Columbia University, 538 West 120th Street, New York, 
New York 10027, USA
}
\date{\today }

\begin{abstract}
The possibility  is investigated that competition between fluctuations at different symmetry-related ordering wave vectors may affect the  quantum phase transition between  a fermi liquid and  a longitudinal spin  density wave state, in particular giving rise to an intermediate  'nematic' state with broken rotational symmetry but unbroken translational symmetry.   At the marginal dimension  the nematic transition is found to be preempted by a first order transition but a weak symmetry breaking field  restores  a second order  magnetic transition with an intermediate regime in which correlations substantially enhance the broken rotational symmetry.  Comparison to recent experiments is made. 

\end{abstract}

\pacs{71.10.Hf, 75.30.Fv, 05.30.Rt, 71.27.+a}

\maketitle


 'Stripe' spin density wave order occurs in many high-$T_c$ cuprate materials \cite{Tranquada96,Ando02,Doiron07}.  A spin density wave is a longitudinal modulation of the spin density ${\vec S}({\bf R})$ characterized by a wavevector ${\bf Q}$ giving the periodicity of the spin modulation. In a 'stripe', $2{\bf Q}$ is not a reciprocal lattice vector so the magnitude of the spin, as well as its direction,  varies from lattice site to lattice site. If the underlying lattice has sufficient symmetry, stripe ordering may occur at  one of several inequivalent wavevectors ${\bf Q}_a$. In the hole-doped high-$T_c$ cuprates  the important physics is two dimensional and the lattice has (to a good approximation) square symmetry.  For dopings greater than about $x=0.05$ and less than a material-dependent number ranging from $0.08$ (in $YBa_2Cu_3O_{6+x}$ \cite{Hinkov08}) to $0.24$ (in $Nd_{2-x}Sr_xCuO_4$ \cite{Ichikawa00,Daou09a,Taillefer09}) order is believed to occur at one of the two wave vectors ${\bf Q}_x=(\pi-\delta,\pi)$ or ${\bf Q}_y=(\pi,\pi-\delta)$ with $\delta$ doping-dependent and generically nonzero  \cite{Tranquada96,Ando02,Hinkov08,Daou09b,Taillefer09,Fradkin09}.  
  
In a stripe state the expectation value of the  spin density at position ${\bf R}$, ${\vec S}({\bf R})$,   may be written
\begin{equation}
\left<{\vec S}({\bf R})\right>={\vec A}_a\hspace{.025in}\cos\left({\bf Q}_a\cdot {\bf R}+\theta_a\right)
\label{stripedef}
\end{equation}
The state defined by Eq \ref{stripedef} breaks  spin rotation, lattice translation and lattice rotation symmetries.  A phase characterized by a nonvanishing $<{\vec S}({\bf R})>$ is also characterized by nonvanishing 
\begin{eqnarray}
\left< {\cal T}({\bf R})\right>&\equiv& \left< {\vec S}({\bf R})\cdot {\vec S}({\bf R})\right>-\left<\left< {\vec S}({\bf R})\cdot {\vec S}({\bf R})\right>\right>
\nonumber \\
&\sim& \cos\left(2{\bf Q}_a\cdot {\bf R}+2\theta_a\right) 
\label{Tdef}
\end{eqnarray}
where the double-bracket indicates also an average over position. ${\cal T}$ is spin rotation invariant but breaks lattice translation and rotation symmetry if $2{\bf Q}$ is not a reciprocal lattice vector. (One may also consider bond order  involving  $<{\vec S}({\bf R})\cdot {\vec S}({\bf R}^{'}\neq {\bf R})>$ but this  will not be important here). ${\cal T}$  couples linearly  to lattice distortions and the electronic charge density, so is observable in scattering measurements \cite{Tranquada96,Fradkin09,Kivelson98} and is sometimes referred to as 'charge order'. 

The 'stripe' state is also characterized by nonvanishing 
\begin{equation}
\left<\eta({\bf R})\right>=\left<\left<{\vec S}_{Q_x}(R)\cdot {\vec S}_{Q_x}(R)-{\vec S}_{Q_y}(R)\cdot {\vec S}_{Q_y}(R)\right>\right>
\label{Rdef}
\end{equation}
where $S_{Q_a}$ indicates spin fluctuations with wave vectors near $Q_a$.  $<\eta>$ is invariant under spin rotations and lattice translations but breaks the discrete lattice rotation symmetry and may be referred to  as a nematic order parameter \cite{Kivelson98,Fradkin09}.  

The three broken symmetries may be restored at separate transitions \cite{Kivelson98,Fradkin09}.  (Very similar phenomena are well understood in the  classical physics context of smetic and nematic liquid crystals \cite{Chaikin95}).  If spin order is destroyed by fluctuations in the direction of ${\vec A}$ (as would happen in a two dimensional model with Heisenberg symmetry at any $T>0$),  $<T>$ and $<\eta>$ may be expected to  remain non-zero.  If long ranged order in $T$ is destroyed by fluctuations in  $\theta$,  $<\eta>$ may remain nonvanishing. In physical terms, the state with $<{\vec S}>=<{\cal T}>=0$ but $<\eta>\neq0$ has the property that   fluctuations around one of the ${\bf Q}_a$ are larger than fluctuations around the other possible ${\bf Q}_{b\neq a}$. 

Experimental evidence suggests that this sequence of transitions indeed occurs in some high $T_c$ compounds.  Many measurements  \cite{Tranquada96,Fradkin09} indicate that in underdoped cuprates the ground state (if superconductivity is suppressed) is characterized by magnetic scattering at the wavevectors  ${\bf Q}_{x,y}$ and  $2{\bf Q}_{x,y}$  but not at ${\bf Q_x}\pm {\bf Q_y}$, implying that the scattering signal arises from a superposition of domains with order at either  ${\bf Q}_x$ or ${\bf Q}_y$. As temperature is raised above an ordered state the Bragg scattering at ${\bf Q}$ vanishes first, leaving an intermediate state with Bragg scattering only at ${\bf 2Q}_x$ and  ${\bf 2Q}_y$   \cite{Tranquada96}.   Recent neutron scattering measurements \cite{Hinkov08} on a monodomain sample of $YBa_2Cu_3O_{6.45}$  indicate  a wide temperature regime where there is  order neither at ${\bf Q}_{x,y}$ nor at $2{\bf Q}_{x,y}$  but where  the fluctuations associated with ordering wavevector ${\bf Q}_x$ have much longer spatial range and stronger temperature dependence  than the fluctuations associated with potential ordering wave vector ${\bf Q}_y$.  Transport measurements have detected rotational symmetry breaking \cite{Ando02} and, recently,   an enhancement of the Nernst effect in this temperature regime has been reported\cite{Daou09b,Choiniere09}, also consistent \cite{Hackl09} with an intermediate nematic phase. Similar transport behavior in the $(Nd/Sr)_2CuO_4$ family of materials  has also been interpreted in terms of a nematic phase or regime \cite{Daou09a}.  It is however important to note that the crystal structure of both of these materials is such that a $CuO_2$ plane is orthorhombically distorted so it may be more appropriate to describe the observed 'nematic' regime  as being characterized by a  strong and strongly doping and temperature-dependent enhancement of a pre-existing anisotropy. 

Closely related issues have been discussed in the context of the pnictide materials \cite{Fang08,Qi09} where $2{\bf Q}$ is a reciprocal lattice vector \cite{Cruz08} so only the spin and nematic orders are relevant. Also in pnictides a strong coupling to lattice distortions believed to be important. 

These and related experiments have focussed theoretical attention on  'nematicity'.  A number of works consider nematic phases which are taken to be conceptually independent of any density wave ordering\cite{Halboth00,Yamase00,Lawler07,Kim07,Huh08,Sun08,Hackl09,Gull09}. This paper considers the density wave instability as primary with the nematic phase arising from it. The  physical idea is straightforward: in a stripe situation, density wave ordering at one possible ordering  wavevector ${\bf Q}_a$ must act to suppress density wave ordering at the other possible wavevectors ${\bf Q}_{b\neq a}$. Thus, as a putative  'stripe' quantum critical point is approached  competition between  fluctuations at different wavevectors may drive a 'nematic' transition at which the system chooses one wavevector at which the fluctuations will become critical, while fluctuations at the other wavevectors remain massive.  Alternative possibilities are that the competition is important only for selecting the relevant state inside the density wave ordered phase, or that   competition between fluctuations may drive the transition first order. In renormalization group language  the question is  whether there is a relevant operator  at the stripe critical point and, if so, does it imply a  flow to a new 'nematic' critical point or a runaway flow indicating a first order transition.  

\begin{figure}[t]
\includegraphics[width=0.8\columnwidth]{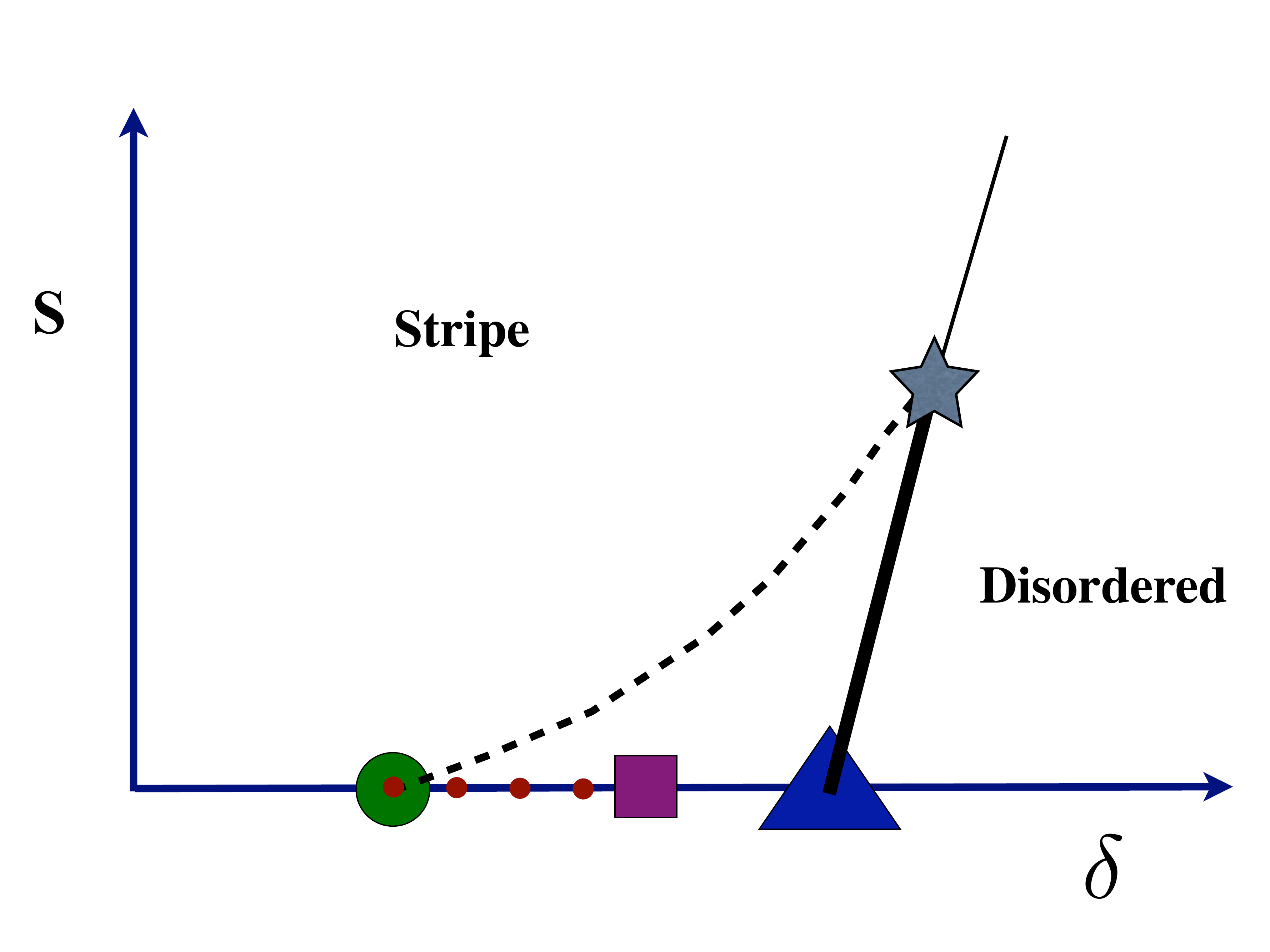}
\caption{Zero temperature phase diagram of two dimensional stripe ordering model in plane of doping $\delta$ and tetragonal symmetry breaking parameter $S$.  Solid lines: phase transition between paramagnetic metal ('disordered') and density wave ("stripe") state. Star (light blue on-line): tricritical point separating first order (heavy line) and second order (light line) phase transitions. Dashed line (black on-line);  putative second order density wave transition preempted by first order transition.   Small filled circles (red on-line): paramagnetic nematic phase, defined only at $S=0$ and separated from density wave and disordered phases by second order phase transition points marked respectively by large circle (green on-line) and square (purple on-line). Nematic phase and associated transitions are  also  preempted by  first order transition }
\label{phasediagram}
\end{figure}

This paper approaches the physics in terms of a $T=0$   instability of  a disordered fermi-liquid  phase using the standard 'Hertz' model of a density wave transition in a two-dimensional fermi liquid \cite{Hertz76,Millis93}.  The main finding is  the phase diagram depicted in Fig \ref{phasediagram}: for lattices with square symmetry the quantum 'nematic' transition is typically preempted by a strongly first-order transition directly to a density wave state; however, a weak explicit symmetry breaking restores a continuous transition.     
 
Incommensurate density wave transitions  are the subject of an extensive literature \cite{Abanov04,Fawcett94},  but the issue of interest here has been less studied. Physics similar to that of interest here has been explored in the context of classical spin models for cuprates \cite{Fang06} and pnictides \cite{Fang08}. Also, although their main focus was on transitions between nematic and density wave ordered states, Sun and co-workers observed \cite{Sun08} that the basic fermi liquid to  density wave transition would likely  be first order.  DePrato, Pelisseto and Vicari \cite{Prato06} used  renormalization group techniques to classify the quantum critical  fixed points of a model of 'stripe' quantum criticality involving  undamped spin excitations in two spatial dimensions, in particular identifying and analysing  regimes of stable second order transitions. At these transitions there would be no intermediate nematic phase separating the density wave and disordered state.  

Pelissetto, Sachdev and Vicari \cite{Pelissetto08} studied a spin density wave transition occurring inside a d-wave superconducting state. The most relevant perturbation involved coupling of nodal fermions to a  nematic order parameter derived from the spin fluctuations in a manner very similar to  what is considered here; however again the density wave transition was not preempted by a nematic one.  Qi and Xu \cite{Qi09}  studied a spin-fermion model, finding a runaway flow also indicating first order transitions, but only at an exponentially long length scale.

A theory of the disordered spin density wave may be obtained from Eq \ref{stripedef} by regarding ${\vec A}$ and $\theta$ as slowly fluctuating quantities. We specialize to the two dimensional square (or rectangular) lattice.   Because we are interested in the transition from a fermi liquid where all spin amplitudes are small we combine $A_a$ and $\theta_a$ into a new complex field ${\vec \psi}_a(r)={\vec A}_a(r)e^{i\theta_a(r)}$ which we assume  is described by the action $S=S_{dyn}+S_{static}$ with 
\begin{eqnarray}
S_{static}&=&\int d^2r d\tau \sum_{a=x,y}\left(\frac{1}{2}\left|\nabla {\vec \psi}_a\right| ^2+\frac{1}{2}\delta_a\left|{\vec \psi}_a^2\right|\right)
\label{Sstatic}
\\
&+&\int d^2r d\tau \frac{u+v}{8}\left(\left|{\vec \psi}_x^*\cdot {\vec \psi}_x\right|+\left|{\vec \psi}_y^*\cdot {\vec \psi}_y\right|\right) ^2
\nonumber\\
&&\hspace{.2in}+\frac{u-v}{8}\left(\left|{\vec \psi}_x^*\cdot {\vec \psi}_x\right|-\left|{\vec \psi}_y^*\cdot {\vec \psi}_y\right|\right) ^2
\nonumber
\end{eqnarray}

The $\delta_a$ are control parameters (for example, doping) which tune the system through the magnetic quantum critical point and  we have allowed for the possibility that deviations from tetragonal symmetry favor ordering in one direction rather than another.  Because the experimental evidence in high-$T_c$ materials indicates that spin fluctuations are stronlgy peaked near discrete momentum values we do not need to consider the possibility of a continuous rotation of the wavevector and any potential nematic phase would have a strong Ising anisotropy. 

Of the six possible quartic nonlinearities (see \cite{Prato06} for a complete list) Eq \ref{Sstatic} includes  only the two which are important for the present purpose, neglecting terms which favor spiral and other non-stripe states or renormalize the basic stiffness against large amplitudes of the fields.   The crucial term is the third one, which quantifies the extent to which orders in the $x$ and $y$ directions compete with each other.    The relevant case is $v>u$, so that fluctuations compete.  On the mean field level stability of this theory requires that $u>0$ and $v>-u$.  

We assume standard overdamped dynamics.  For  ease of writing we present $S_{dyn}$ in frequency space:
\begin{equation}
S_{dyn}=\frac{1}{2}\sum_{a=x,y}T\sum_n\int d^2r\frac{|\Omega_n|}{\Gamma}|{\vec \psi}^a\cdot{\vec \psi}^a|
\label{Sdyn}
\end{equation}
The theory requires an ultraviolet cutoff. We measure energy in units of $\Gamma$ and impose a hard cutoff, eliminating all processes for which  $|\Omega|/\Gamma+k^2>\Lambda$.  Qi and Xu \cite{Qi09} studied essentially this model, but with an additional $({\vec \psi}_1\cdot {\vec\psi}_2)^2$ coupling.

As defined the upper critical dimension of the model is $d=2$ and the physics may be studied by a renormalization-group analysis. The required beta functions are given in Eq (3.1) of \cite{Prato06}. It is useful define new variables $g$ and $\phi$ by $u=g \cos\phi$, $v=g\sin\phi$ which flow according to (here the dot denotes changes with renormalization group cutoff parameter)
\begin{eqnarray}
\dot{g}&=&-\left(7\cos^3\phi+11\cos\phi \sin^2\phi+2\sin^3\phi\right)g^2
\label{gflow}\\
\dot{\phi}&=&\left(3\sin^3\phi-2\sin^2\phi \cos\phi -\sin\phi \cos^2\phi\right)g
\label{phiflow}
\end{eqnarray}

Because $\dot{g} \sim g^2$ while  $\dot{\phi}\sim g$, $\phi$ flows much more raipdly than $g$ and the content of the theory may be understood from a constant $g$.  In Eq \ref{phiflow} the angle $\phi/4$ (corresponding to $u=v$) is a separatrix. For $\phi<\pi/4$ the flow is towards $\phi=0$, but in the $\phi>\pi/4$  case relevant to stripe physics the flow is towards a fixed point value which is close to  $\pi$.  Noting that $u$ turns negative at  $\phi=\pi/2$ we see that  the renormalization group analysis indicates that when the flow passes this point  the two dimensional stripe fixed point becomes unstable towards a first order transition.  The basic  conclusion is perhaps not surprising: the model of two coupled order parameter fields is a textbook example of a runaway flow leading to a first order transition \cite{Rudnick78,Cardy96}.  De Prato et al \cite{Prato06} observed that their more general renormalization group equations had only unstable fixed points near the marginal dimension and Sun et. al. noted that multicritical points of this type tend to be unapproachable due to the presence of runaway flows. 

Qi and Xu \cite{Qi09} similarly noted the possibility of a runaway flow. The equations of \cite{Qi09} involve three couplings and are thus more complicated to solve. A numerical solution was presented  which indicated a runaway flow, albeit beginning at an exponentially low scale, whereas what is found here is a first order transition at  a scale which is not, in general, exponentially small.  

The first order nature of the transition arises from competition between fluctuations associated with the two ordering wave vectors. An explicit symmetry breaking term (arising e.g. from the chains in $YBCO$) would grow  under renormalization and if it became large enough, would quench the fluctuations at one of the two wave vectors, thereby permitting a continuous behavior.  To understand the energy scales involved in this scenario we consider a self-consistent one-loop analysis, which while less rigorous than a renormalization group treatment has a transparent physical interpretation and allows for straightforward estimations of energy scales. 

To implement the self consistent one loop theory we write the model as a functional integral and decouple the nonlinearities $|\psi_x|^2+|\psi_y|^2$ and $|\psi_x|^2-|\psi_y|^2$ by Hubbard-Stratonovich fields $i\lambda$ and $\eta$ respectively and then integrate over the $\psi$ fields obtaining the action  
\begin{eqnarray}
S[\lambda,\eta]&=&\frac{\lambda^2}{2(v+u)}+\frac{\eta^2}{2(v-u)} +\frac{3}{2}Tr \ln [ \Pi_0+\delta_x+i\lambda+\eta] 
\nonumber \\
&+&
\frac{3}{2}Tr \ln[\Pi_0+\delta_y+i\lambda-\eta] 
\label{actionpm}
\end{eqnarray}    
with $\Pi_0=|\Omega|+k^2$. Eq \ref{actionpm} is written for the paramagnetic phase; the factor of $3$ is the spin degeneracy.  
Mean field theory corresponds to finding the $\lambda^*$ and $\eta^*$ which extremize $S$. The extremal values of $\lambda$ are imaginary; we write $i\lambda={\bar \delta}-r$ with ${\bar \delta}=(\delta_x+\delta_y)/2$ and introduce $\Delta=(\delta_x-\delta_y)/2$ which parametrizes any explicit breaking of tetragonal symmetry. The correlation length $\xi_{x,y}$ for spin fluctuations around the wavevectors ${\bf Q}_{x,y}$ is $\xi_{x,y}^{-2}=r\pm(\eta+\Delta)$.  In the paramagnetic phase $ \xi^{-2}_a >0$. If one of the fields, say $\psi_y$, orders, then the Heisenberg symmetry ensures that  the two transverse components are gapless ($\xi_{\perp,y}^{-2}=0$) while the longitudinal component has a correlation length determined by the magnetization $m$, $(\xi_{\parallel,y}^{-2}=m^2$) so that the term $\frac{3}{2}Tr \ln[\Pi_0+\delta_y+i\lambda-\eta] \rightarrow Tr \ln[\Pi_0]+\frac{1}{2}Tr \ln[\Pi_0+m^2]$.  

Fang et. al. \cite{Fang08} presented a large-N analysis of a classical spin model which leads to a theory very similar to that defined by Eq \ref{actionpm} if ${\bar \delta}$ is chosen to be deep in the ordered phase so quantum fluctuations are unimportant.

Defining $\delta_{crit}$ to be the value at which $S$ is extremized at  $\Delta=r_x=r_y=0$, redefining   $\delta_{x,y}$ as the difference from  $\delta_{crit}$, introducing  ${\bar u}({\bar v})=3u(v)/4\pi^2$ and explicitly carrying out the minimization in the paramagnetic phase at $T=0$ we obtain:
\begin{eqnarray}
\delta_x&=&r_x\left(1+{\bar u}\ln\frac{1}{r_x}\right)+{\bar v}r_y\ln \frac{1}{r_y}
\label{dx}
\\
\delta_y&=&r_y\left(1+{\bar u}\ln\frac{1}{r_y}\right)+{\bar v}r_x\ln \frac{1}{r_x}
\label{dy}
\end{eqnarray}

In the   tetragonal symmetry case $\Delta=0$ and at  ${\bar \delta}>0$ Eqs \ref{dx},\ref{dy} admit an isotropic solution $r_x=r_y>0$ corresponding to the conventional paramagnetic phase.  However, for  $v>u$ one finds that as $\delta$ is decreased below a critical value  $\delta_{nem}>0$ the isotropic solution undergoes a bifurcation to a solution with $r_x\neq r_y$. (Fang et. al. \cite{Fang08} found a very similar  transition, in their case thermally driven). This is a nematic phase: the nematic order parameter is $\eta\sim<|\psi_x|^2>-<|\psi_y|^2>$. Writing $r_x=r+\eta/2$, $r_y=r-\eta/2$ and linearizing in $\eta$ we find 
\begin{equation}
\delta_{nem}=\frac{2{\bar v}+{\bar v}^2-{\bar u}^2}{{\bar v}-{\bar u}}e^{-1-\frac{1}{{\bar v}-{\bar u}}}
\label{deltanem}
\end{equation}

However, within mean field theory this critical point is typically preempted by a first order transition to a state with long-ranged stripe order.  The first order transition manifests itself as a failure of numerical routines to find a solution to Eqs \ref{dx},\ref{dy} as $\delta$ is decreased below a spinodal value greater than $\delta_{nem}$, and may also be seen more directly.  

We have computed the energy of a magnetized state by generalizing  Eq \ref{actionpm}  as described above. We find that at $\delta=\delta_{nem}$ an ordered state with $r=\pm \eta$ and $m>0$  exists and has lower energy than the paramagnetic state, provided that ${\bar v}$ is not too large. Thus at some $\delta>\delta_{nem}$ the system will jump from the isotropic paramagnetic phase to a phase with long ranged order. The largest $\delta$ at which an ordered phase may be sustained may be ascertained from the $\delta$ at which there is a  solution to   Eqs \ref{dx},\ref{dy} with $r_y=0$ and $r_x>0$. This  $\delta=\delta_{1^{st}-order}$ is 
\begin{equation}
\delta_{1^{st}-order}=\frac{{\bar v}e^{-\frac{1}{{\bar v}-{\bar u}}}}{{\bar v}-{\bar u}}=\frac{e}{2+{\bar v}-\frac{{\bar u}^2}{\bar v}}\delta_{nem}
\label{deltaord}
\end{equation}
By comparison of energies we find that if $\delta_{1^{st}-order}>\delta_{nem}$ the nematic transition is preempted by a first order transition.  As the second equality of Eq \ref{deltaord} shows, for ${\bar v}>v_c(u)$ with $v_c(u=0)=e-2\approx 0.718..$, $\delta_{nem}>\delta_{order}$ so that a second order transition can exist at large $v$.  However, in the large ${\bar v}$ regime in which the transition (within the present theory) is second order, substitution into the defining equations shows that  the renormalized mass $r$ is of the order of the cutoff, indicating that the second order transition occurs in a regime beyond the range of validity of the critical theory. We interpret this result as meaning  that  the nematicity arising in the large $v$ case is an intrinsic phenomenon arising from short length scale physics and not directly related to the singular magnetic critical fluctuations.

\begin{figure}[t]
\includegraphics[width=0.8\columnwidth]{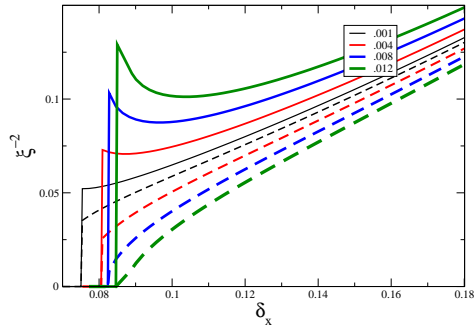}
\caption{Dependence on control parameter $\delta$ of inverse correlation length parameters for $x$ (solid line) and $y$ (dashed line) wavevectors for different levels of explicit symmetry breaking $\delta_x=\delta_y+\Delta$ with ansiotropy parameters $\Delta$ as indicated and interactions ${\bar u}=0.3$ and ${\bar v}=0.6$.}
\label{aniso}
\end{figure}

We are now in a position to consider the effects of an explicit  breaking of the $C_4$ lattice rotational symmetry, such as occurs in $YBa_2Cu_3O_{6+x}$ and  $Nd_{2-x}Sr_xCuO_4$. This acts analogously to a magnetic field at a ferromagnetic transition and, if large enough, will  convert the first order transition to a continuous one. The small value of $(\delta_{1^{st}-order}-\delta_{nem})/\delta_{nem}$ suggests that the symmetry breaking field need not be large.  Fig \ref{aniso} presents results obtained by solving Eqs \ref{dx},\ref{dy}  for $\delta_x=\delta_y+\Delta$ for a representative choice of parameters ${\bar u}=0.3$ and ${\bar v}=0.6$ (implying $\delta_{1^{st} order}=.071$ and $\delta_{nem}=0.0328$). and several levels of anisotropy. One sees that for these parameters  anisotropy greater than about $.01 \sim (\delta_{1^{st} order}-\delta_{nem})/4$  restores a continuous behavior.

Extending the results  to temperatures $T>0$ is complicated by issues including the difficulty of treating the classical transitions  below their upper critical dimension, the interplay between 'nematic' and $2{\bf Q}$ (``charge") order and the possible first-order nature of the transitions. A detailed discussion will be given elsewhere but a few remarks are in order here.   

One should distinguish two situations: either the model has an intrinsic tendency towards nematic behavior, not directly related to the stripe criticality (in the model defined by  Eq \ref{Sstatic} this would occur if $v$ is large, both absolutely and relative to $u$)  or the first order transition preempts the $T=0$ nematic phase.  Models with 'intrinsic' nematicity have been extensively discussed elsewhere \cite{Halboth00,Yamase00,Lawler07,Kim07,Huh08,Sun08,Hackl09,Gull09} and will not be considered here.  In the case where the $T=0$ nematic phase is preempted by a first order transition there remains the possibility of nematic behavior at $T>0$. 

Indeed the model defined in Eq \ref{Sstatic}  trivially exhibits  nematic behavior if parameters are chosen so that the model is in the magnetically ordered state at temperature $T=0$ and if charge order is not important (for example because $2{\bf Q}$ is a reciprocal lattice vector). The ordered state must select one of the two wavevectors, say ${\bf Q}_x$.  The Heisenberg symmetry and two dimensionality then implies that if temperature is slightly raised above the ordered state, fluctuations near ${\bf Q}_x$ have a correlation length of the 'renormalized classical' form $\xi \sim exp[\rho_s/T]$ while fluctuations near ${\bf Q}_y$ have a relatively short and weakly temperature dependent correlation length, so the resulting state is a 'nematic'.  The question then is whether this 'thermal spin nematic' behavior vanishes via a first or a second order transition. If a weak interlayer coupling is added to the model one may ask if the nematic behavior survives at temperatures higher than the three dimensional ordering temperature (this issue was addressed in the classical model of Ref [\onlinecite{Fang08}]).  If charge order is also important, one may ask if the nematic behavior vanishes at the charge ordering temperature or at a higher temperature, and what are the orders of the transitions.

Clearly a transition which is  first order at $T=0$  remains first order as the phase boundary is extended to $T>0$, at least within some distance of the zero temperature transition, so an extension of the present results to $T>0$ would suggest a first order thermal transition, at least for dopings near where the $T=0$ ordered state vanishes, but it is not ruled out that the transition may become second order as doping is reduced deeper into the ordered phase. 

The self-consistent one-loop approximation employed above leads to  a reentrant behavior.  As $T$ is increased from zero  the first order transition line extends to the larger $\delta$ side of the critical point and the transition is from  an isotropic paramagnetic phase at low $T$ to a higher $T$ 'nematic renormalized classical' phase  in which  the spin fluctuations near one of the two ordering wave vectors  have a very long and rapidly temperature-dependent correlation length while those near  the other  have a correlation length which is relatively short and weakly temperature dependent.   Whether  the 'nematic renormalized classical' phase  is also characterized by long ranged $2{\bf Q}$ charge order depends on details.  As the temperature is further increased the charge order (if any)  disappears and there is a second transition (typically also first-order within this approximation) to a fully isotropic phase.  The first order transition persists as parameters are tuned to move deeper into the insulating phase. While the reentrant behavior seems unphysical, and is likely to be an artifact of the approximation, the qualitative result of a range of zero temperature parameters above which the thermal transition is first order is robust.  

The classical-spin results of \cite{Fang08} provide an interesting perspective on this issue.  As remarked above, this model  is in essence the self-consistent one-loop approximation to the  classical limit of the  model defined in Eq \ref{actionpm} for $\delta$ deep within the $T=0$ ordered phase and $2{\bf Q}$ a reciprocal lattice vector.   The authors of Ref [\onlinecite{Fang08}] considered a regime in which the physics was controlled by the 'renormalized classical' divergence of the correlation length and moreover chose parameters corresponding to an extremely weak tendency towards nematic ordering. Ref \cite{Fang08} defines a dimensionless quantity  ${\tilde K}/{\tilde J}_2$, which is essentially $(v-u)/(v+u) $ in the  notation of this paper. This was taken to have the very small value $7.5\times 10^{-3}$. For this parameter value Ref [\onlinecite{Fang08}]  reported a second order 'nematic' transition which occurred at very low temperatures deep in the 'renormalized classical' regime and  was not preempted by a first order transition.   Ref [\onlinecite{Fang08}] did not report results for larger ${\tilde K}$, but solving their equations indicates that as  ${\tilde K}$ is increased to a value $\gtrsim 0.075{\tilde J}_2$ the thermally driven transition again becomes first order.  Now $4\pi {\tilde J}_2$ is essentially the spin stiffness of the magnetically ordered state which may be thought to grow with distance into the ordered phase so if one imagines that ${\tilde K}\sim (v-u)$ is fixed to a value which is not too large, the increase in the magnetic spin stiffness  distance into the ordered phase may eventually drive the model into the second order transition regime.

Of course, the reliability of the self-consistent one-loop approximation may be questioned, especially for thermal phenomena. The analysis of \cite{Millis93} indicates that this analysis is essentially equivalent to a renormalization group analysis for models above the upper critical dimension, but for models which are below the upper critical dimension this is of course not the case. The self-consistent one loop approximation has the exponents of the Gaussian model, and for this reason may overestimate the tendency towards nematic order. For example, one may define a 'nematic susceptibility' via the correlation function $<(\psi_x^2-\psi_y^2)^2>$. This correlator is closely related to the energy correlator that defines the specific heat exponent. The gaussian model dramatically overestimates the divergence of the specific heat near $2D$ and $3D$ classical transitions and may well similarly overestimate the divergence of the 'nematic susceptibility'.

Despite these difficulties it is interesting to relate the picture presented here to data. The essential point is that in models in which 'nematicity' is derived from competition between density wave ordering at several wavevectors, the nematic phase may be pre-empted by a first order transition, but if the lattice rotational symmetry is explicitly broken a more continuous behavior may be restored.   There is no direct evidence that  stripe order  terminates at a first order  quantum critical point in any cuprate, although we note that phase coexistence, a typical consequence of first order phase boundaries, is common in cuprate materials. The strongest evidence in favor of a nematic regime comes from the $YBCO$ and $(Nd/Sr)_2CuO_4$ families of materials \cite{Hinkov08,Daou09b,Choiniere09} where the lattice symmetry is explicitly broken. It is very tempting to argue that the symmetry breaking pushes  the system into the continuous transition regime identified above.   It may also be worthwhile to reexamine the data to determine if a local lattice distortion (favoring one or the other wave vector) is present or if a hysteresis has been overlooked. 

In summary, this paper has posed the question of the existence of an intermediate nematic phase separating a stripe and a fermi liquid phase in terms of fluctuation corrections to a putative 'stripe' quantum critical point.  The physics that can lead to a nematic phase also leads naturally to a fluctuation-driven first order transition, which near the quantum critical point was found to preempt the nematic phase. The first order transition is not inevitable. If parameters were tuned so that the nematic transition occurs 'far' from the putative stripe quantum critical point so that the 'nematicity' is an intrinsic effect and not driven by critical density wave fluctuations, then the considerations of this paper are not relevant.  

It was found that the first order transition could be converted to second order by a quite small anisotropy. The analysis relied on approximations including the self-consistent one-loop theory (which is uncontrolled) and the Hertz quantum critical theory (which is subject to corrections whose nature remains incompletely understood \cite{Abanov04} ) but the  first order behavior discussed here follows from relatively general considerations and seems likely to be robust. In systems where nematic behavior was found a re-examination of experimental data for signatures of first-order transitions (for example,  hysteresis and phase coexistence) may be worthwhile.  Extension of the results presented here to $T>0$, to include coupling to the lattice, and to other situations, such as the metamagnetic transition in  $Sr_3Ru_2O_7$, would be of interest. 

{\it Acknowledgements:} I thank    E. Fradkin, E. Kim, M. Lawler, T. Lubensky, M. Norman,   S. Sachdev, J.Schmalian,   L. Taillefer and C. Xu for helpful discussions and the  New York University Department of Physics for hospitality. This work was supported by NSF-DMR-0705847 and benefitted in an essential way from interactions with members of the CIFAR quantum materials program.

\end{document}